\documentclass[aps,prd,amsmath,floats,floatfix,superscriptaddress,nofootinbib,showpacs,onecolumn]{revtex4}

\usepackage{amssymb}
\usepackage{amsmath}
\usepackage{verbatim}
\usepackage{mathrsfs}
\usepackage{amsfonts}
\usepackage{latexsym}
\usepackage{epsfig}
\usepackage{color}
\usepackage{graphicx}
\usepackage{units}
\usepackage{overpic}
\usepackage[hidelinks]{hyperref}
\usepackage{afterpage}
\usepackage{float}
\usepackage{mathtools}
\usepackage{xcolor}
\hypersetup{
    colorlinks,
    linkcolor={red!50!black},
    citecolor={blue!50!black},
    urlcolor={blue!80!black}
}

\usepackage{booktabs, multirow} 
\usepackage{soul}
\usepackage{changepage,threeparttable} 
\begin{document}


\definecolor{orange}{rgb}{0.9,0.45,0} 
\definecolor{applegreen}{rgb}{0.55, 0.71, 0.0}
\definecolor{blue}{rgb}{0.0,0.0,1.0}

\newcommand{\juanc}[1]{{\textcolor{green}{[JCD: #1]}}}


\title{Hayward Boson Stars}

\author{Sebastian S. Chicaiza-Medina}
\affiliation{Escuela Superior Politecnica de Chimborazo (ESPOCH), Riobamba 060155, Ecuador.}
\affiliation{Instituto de Ciencias F\'isicas, Universidad Nacional Aut\'onoma 
de M\'exico, Apartado Postal 48-3, 62251, Cuernavaca, Morelos, M\'exico.}

\author{Juan Carlos Degollado} 
\affiliation{Instituto de Ciencias F\'isicas, Universidad Nacional Aut\'onoma 
de M\'exico, Apartado Postal 48-3, 62251, Cuernavaca, Morelos, M\'exico}


\date{\today}

\begin{abstract}

We examined the Einstein-Klein-Gordon system coupled to a nonlinear electrodynamics framework that asymptotically supports a Hayward spacetime.
We explored the solution space of a static spherically symmetric, complex scalar field 
minimally coupled to the gravitational field  known as Hayward Boson stars originally studied by Yue and Wang in \cite{Yue:2023sep}.
We construct families of Hayward boson stars in the ground state, for different values of the charge parameter $Q$, and different values of the central scalar field. 
One of the main results of our analysis is the fact that
the existence of boson stars is guaranteed once the condition $\frac{\sqrt{\beta}}{Q}> 1.49661$, where $\beta$ is a parameter of the electrodynamic theory, is fulfilled.
When this condition is not satisfied the electrovacum spacetime contains at least one horizon and the spacetime can not support a scalar field configuration. 
\end{abstract}




\maketitle


\section{Introduction}
\label{sec:intro}

Boson stars (BS) are self-gravitating, regular and asymptotically flat configurations of massive bosonic fields (either scalar or Proca fields)
whose gravitational collapse is prevented by the Heisenberg uncertainty principle \cite{Jetzer:1991jr,Liebling:2012fv,Schunck:2003kk} . These gravitationally bound objects arise as solutions in 
General Relativity (GR) coupled to massive bosonic fields with a continuous symmetry \cite{Ruffini.187.1767,Kaup:1968zz}. The existence of this symmetry allows us to build a stationary stress-energy tensor that via Einstein's equations produces a stationary spacetime. The simplest model of boson stars is the one that considers spherical symmetry and a potential with a mass term, these are also called mini-boson stars \cite{Colpi:1986ye,Schunck:1999zu,Ho:1999hs}.

Generalizations of BSs by including couplings to electromagnetic fields have been extensively studied, some examples are found in Refs. \cite{Jetzer:1989av,Jetzer:1989us,Pugliese:2013gsa,Kan:2017rqk,Visinelli:2021uve,Kleihaus:2009kr}. These studies enriched the range of possible solutions in scalar field theories allowing for a more comprehensive understanding of how scalar fields interact with fundamental forces, particularly gravity and electromagnetism.

Despite lacking an event horizon, the external spacetime geometry of a BS can closely mimic that of black holes (BH), particularly in terms of mass, compactness,
and gravitational lensing properties \cite{Olivares:2018abq,Grandclement:2014msa}. As a result, BSs are capable of reproducing several observational signatures typically associated with black holes,
such as the bending of light\cite{Grandclement:2016eng}, quasi-normal mode spectra \cite{Yoshida:1994xi,Macedo:2013jja}, and accretion disk behavior \cite{Torres:2002td,Guzman:2005bs,Guzman:2009zz}.

Although classical BHs are well described by GR, they present a singularity where the theory is unable to give a complete description \cite{Penrose:1964wq,Wald:1984rg}.
Many efforts have been made in order to alleviate the problem of singularities in GR. The seminal work of Bardeen on a regular black hole \cite{Bardeen,Ayon-Beato:2000mjt,Lan:2023cvz} gave rise to several studies exploring nonsingular solutions within the GR framework, inspiring a broad class of models in which central singularities are avoided through modifications to the matter content. A common path to describe regular BHs
within GR has been considering Non-linear Electrodynamics (NLED) sourcing the spacetime \cite{Ayon-Beato:1998hmi,Ayon-Beato:1999kuh,Bronnikov:2017sgg,Bronnikov:2022ofk}.

The Hayward solution \cite{Hayward:2005gi}
introduces a carefully constructed mass function designed to eliminate the central singularity in such a way that the 
mass function interpolates smoothly between the Schwarzschild behavior at large distances and a de Sitter-like core near the origin.
Later, Fan and Wang showed in \cite{Fan:2016hvf} that the underlying physical mechanism responsible for this regularity can be modeled through a  NLED 
incorporating a magnetic monopole
source in Einstein’s field equations. In this context, magnetic monopole-like configurations arising from the NLED field provide the necessary energy conditions to support a non-singular geometry (see also \cite{Bronnikov:2017tnz}).

The Hayward metric has been used as a prototypical example of a regular BH while remaining asymptotically flat and maintaining some key features expected from a compact object \cite{Cardoso:2019rvt}.

The purpose of this paper is to describe the properties of non-charged boson stars in a NLED model inspired by Hayward spacetime
as done in \cite{Yue:2023sep} where Hayward Boson Stars (HyBS) were introduced for the first time. 

HyBS are obtained
by coupling Einstein gravity to a massive, uncharged scalar field and incorporating the NLED framework proposed by Fan and Wang originally developed to
support the Hayward spacetime.
These configurations represent self-consistent, regular solutions where the interaction between gravitational, scalar, and electromagnetic fields gives rise to relevant structures within the context of modified compact objects.
We consider a non-charged scalar field such that the bosonic component interacts with the electromagnetic component only via the gravitational field.
Although different couplings are possible, such a simple interaction allows us to extract some conclusions regarding the existence and properties of solutions of the system.
In Ref. \cite{Wang:2023tdz} other NLED inspired in the Bardeen spacetime have been used to model Bardeen stars. In that work, the emphasis has been on the limiting behavior of solutions whose 
frequency $\omega\rightarrow 0$, describing \emph{frozen stars} \cite{Chen:2024bfj}. 

In this work, we present sequences of HyBS solutions and describe some of their properties. 
We construct families of HyBSs, concentrating on the ground state configurations for various values of the charge parameter 
$Q$, and different central amplitudes of the scalar field.

One of the key findings of our analysis is that HyBS solutions exist only when the ratio $\frac{\sqrt{\beta}}{Q}$, where $\beta$
is the electrodynamic parameter described below, exceeds a certain critical threshold. This condition defines the region in the parameter space where regular, self-gravitating scalar field configurations can form.

The paper is organized as follows. In section \ref{sec:setup}, we introduce the model of a scalar field in Einstein gravity coupled to a NLED and describe
 the requirements to construct a spherically symmetric object, including boundary conditions.
 In section \ref{sec:results}, we construct a variety of solutions of BSs and compare them with the known Hayward spacetime. We also revisit some of the properties of a Hayward spacetime with no horizons.
By the end of the section, we describe the spacetime of the HyBs.
Finally, in section \ref{sec:conclusions} we provide a brief discussion on the results and draw some conclusions.

\section{Set up}
\label{sec:setup}

We consider the Einstein-electromagnetic-scalar model with the action given by 
\begin{eqnarray}
\label{eq:action}
 \mathcal{S}=\int d^4 x
\sqrt{-g}
\left[
\frac{R}{16\pi}
-\mathcal{L}(F)
-\nabla_\alpha \Psi^*\nabla^\alpha \Psi-U(\left|\Psi\right|)
\right]~,
\end{eqnarray} 
$R$ is the Ricci scalar associated with the
spacetime metric $g_{\mu\nu}$, which has determinant $g$,
 $F_{\alpha\beta} =\partial_\alpha A_\beta - \partial_\beta A_\alpha$ is the Maxwell 2-form, $F=F^{\alpha\beta} F_{\alpha\beta}$, $A_\alpha$ is the gauge 4-potential, $\Psi$ is a complex scalar field, `*' denotes complex conjugate and 
  $ U(|\Psi|)$ denotes the scalar potential, which in this work is taken as $U = \mu^2 |\Psi|^2$. The parameter $\mu$ is the associated mass of the scalar field particle \cite{Alcubierre:2022rgp}.
 
We shall consider the electromagnetic lagrangian $\mathcal{L}(F)$ that was proposed by Fan and Wang \cite{Fan:2016hvf} 
as the one that, coupled with Einstein's gravity, produces a Hayward spacetime and is given by
\begin{eqnarray}
    \mathcal{L}(F) = \frac{3}{\beta\pi} \frac{\left(\beta F\right)^{3/2}}{\left[1+ (\beta F)^{3/4}\right]^2} \; .  
\end{eqnarray}

The Einstein--Maxwell-scalar field equations, 
obtained by varying \eqref{eq:action} with respect to the metric, scalar field and electromagnetic field, are, respectively,
\begin{eqnarray}
\label{Eq:Einstein}
&&
R_{\alpha\beta}-\frac{1}{2}g_{\alpha\beta}R=8 \pi  \left[ T_{\alpha\beta} ^{\rm (EM)} +T_{\alpha\beta}^{(\Psi)} \right] \ , 
\\
\label{eq:kg} 
&&
\nabla_{\alpha}\nabla^{\alpha}\Psi=\frac{d U}{d\left|\Psi\right|^2} \Psi \ , \qquad 
~~\\
\label{eq:max} 
&&
\nabla_{\alpha}\left( \mathcal{L}_F F^{\beta \alpha}\right)= 0 \; ,
\end{eqnarray}  
where  $\mathcal{L}_F = dL/dF$. Notice that, since the scalar field is not charged, the Maxwell euqations Eq. \eqref{eq:max} decouple from the Klein-Gordon equation Eq. \eqref{eq:kg}, this will be important for our further results.

The electromagnetic and scalar components of the energy-momentum tensor are
\begin{eqnarray}
\label{tmunu} 
T_{\alpha\beta}^{\rm (EM)}&=&\mathcal{L}_FF_\alpha{}^\delta F_{\beta\delta}-\frac{1}{4}\mathcal{L}(F)g_{\alpha\beta} \ ,~
\\
\nonumber
T_{\alpha\beta}^{(\Psi)}
&=&
 \nabla_\alpha\Psi^* \nabla_\beta\Psi 
+\nabla_\beta\Psi^* \nabla_\alpha\Psi  
-g_{\alpha\beta}  \left[ \frac{1}{2} g^{\gamma\delta} 
 ( \nabla_\gamma \Psi^* \nabla_\delta\Psi+
\nabla_\delta\Psi^* \nabla_\gamma\Psi) +U(\left|\Psi\right|) \right]
 \ .
\end{eqnarray} 

\subsection{Spherically symmetric spacetime.}

We are interested in static solutions to the Einstein equations thus we take the metric as
\begin{eqnarray}
\label{eq:metric-esf}
    ds^2= - N(r)\sigma(r)^2dt^2 + \frac{dr^2}{N(r)}+  r^2(d\theta^2 +\sin^2\theta d\varphi^2) \; , \qquad N(r) = 1-\frac{2 m(r)}{r} \; ,
\end{eqnarray}
where $N(r)$ and $\sigma(r)$ are functions of the areal radius $r$.\\
For the scalar field we assume a time dependence of the form,
\begin{eqnarray}
\Psi=\psi(r) e^{-i \omega t}\ , 
\label{eq:scal-ans}
\end{eqnarray}
where $\omega$ is the oscillation frequency of the scalar field. 
For the electromagnetic potential we assume the form
\begin{eqnarray}
 A= a(\theta) \;d{\varphi} \ ,
\label{eq:4pot-ans}
\end{eqnarray}
Thus,
the nonvanishing components of the Faraday tensor are
\begin{eqnarray}
F_{\theta\varphi}=-F_{\varphi\theta} = \partial_\theta \,a \; , \quad \Rightarrow \quad F = \frac{2(\partial_\theta \,a)^2}{r^4}\csc^2\theta \; .
\end{eqnarray}
The Maxwell equations \eqref{eq:max} with the ansatz for the four potential \eqref{eq:4pot-ans} 
and the form of the spacetime metric \eqref{eq:metric-esf}
yield a differential equation for $a(\theta)$, 
\begin{eqnarray}\label{eq:a}
\partial_{\theta\theta}\; a-(\cot\theta)\partial_\theta \,a   = 0 \; ,
    \end{eqnarray}
that can be solved directly giving the four potential as
\begin{eqnarray} 
\label{eq:sol-max}
   a(\theta) = Q\cos\theta \; ,
\end{eqnarray}
where $Q$ is a parameter that can be interpreted as the magnetic charge source.

The previous ansatz for the scalar field and the solution \eqref{eq:sol-max} leaves three radial functions to be determined $\sigma(r),m(r),\psi(r)$.      The corresponding field equations, resulting from (\ref{Eq:Einstein},\ref{eq:kg})
read, denoting radial derivatives by ``primes",
\begin{eqnarray}
&&
\label{eq:m}
m'=4\pi  r^2
\left[N\psi'^2+{\mu^2} \psi^2+\frac{w^2}{N\sigma^2}\psi^2
\right] +
\frac{3 r^2}{\beta} \frac{(\beta F)^{3/2}}{\left[1+ (\beta F)^{3/4} \right]^2} \; ,
\\
&&
\label{eq:sigma}
\sigma'=8\pi  r \sigma
\left[
\psi'^2+\frac{w^2\psi^2}{N^2 \sigma^2}
\right] \ ,
\\
&&
\label{eq:psi}
\psi''+
\left(
\frac{2}{r}+\frac{N'}{N}+\frac{\sigma'}{\sigma}
\right)\psi'
+\frac{ \omega^2 \psi}{N^2\sigma^2}
-\frac{\mu^2\psi}{N}
=0~.
\end{eqnarray}
Alternatively, the equation for the mass Eq. \eqref{eq:m}, can be written as
\begin{eqnarray}
\label{eq:m-rho}
m'=4\pi  r^2
\left(\rho_{\Psi}+  \rho_{\rm EM}\right) 
\ ,
\end{eqnarray}
where
\begin{eqnarray}
\label{eq:energy-density}
\rho_{\Psi}= N\psi'^2+{\mu^2} \psi^2+\frac{w^2}{N\sigma^2}\psi^2 \; ,
\qquad
    \rho_{\rm EM} = \frac{3}{4\pi\beta} \frac{(\beta F)^{3/2}}{\left[1+ (\beta F)^{3/4} \right]^2} \; , \qquad F= \frac{2Q^2}{r^4} \; .
\end{eqnarray}
Furthermore, it is convenient for numerical purposes, to write the Klein Gordon equation, Eq. \eqref{eq:psi}, as
\begin{eqnarray}
\label{eq:kg-mol}
    \psi''+ \frac{\psi'}{Nr}\left(1+N-   {8\pi r^2 \mu^2\psi^2}- \frac{6 r^2}{\beta}
    \frac{(\beta F)^{3/2}}{\left[1+ (\beta F)^{3/4} \right]^2}\right)+\frac{ \omega^2 \psi}{N^2\sigma^2}- \frac{\mu^2\psi}{N} =0\; ,
\end{eqnarray}
where we have used Eq. \eqref{eq:sigma} and 
\begin{eqnarray}
N'+r\left[
N\left(  \frac{1}{r^2}+ 8\pi \psi'^2   \right)
+8\pi\left(\mu^2+\frac{\omega^2}{N\sigma^2}\right)\psi^2
-\frac{1}{r^2}
+ \frac{3}{\beta}
    \frac{(\beta F)^{3/2}}{\left[1+ (\beta F)^{3/4} \right]^2}
    \right]=0
\end{eqnarray}
that corresponds to the $t\,t$ component of Einstein's equation.

\subsection{Boundary conditions}

In order to numerically solve the field equations, it is necessary to specify suitable boundary conditions. 
Inner boundary conditions near the origin are found by expanding the fields in a power series, substituting the power series into  (\ref{eq:m}, \ref{eq:sigma},  \ref{eq:kg-mol}), and then equating coefficients of the same powers. 
The corresponding behavior is
\begin{eqnarray}\label{eq:initial}
    m(0) &\sim& \left[\frac{4\pi}{3} \psi_0^2\left(\frac{\omega^2}{\sigma_0^2}+1\right) +\frac{1}{\beta} \right]r^3 \; , \\
    \psi(0)&\sim& \psi_0 \; ,\\
    \sigma(0) &\sim& \sigma_0 + 4\pi \frac{\omega^2}{\sigma_0^2} \psi_0^2 r^2\; .
\end{eqnarray}
which is parameterized in terms of the three unknown constants $\omega$, $\sigma_0$ and the value of the scalar field at the origin $\psi_0$. 
The outer boundary conditions require that the scalar field vanishes $\psi \rightarrow 0$, as $r\rightarrow \infty$ and that the
spacetime is asymptotically Schwarzschild and thus $\sigma\rightarrow 1$. Furthermore, the existence of decaying modes is given by the asymptotic behavior of the scalar field
  \begin{equation}
  \psi(r) \sim e^{-(\sqrt{\mu^{2}-\omega^{2}})r}\quad\text{with}\quad \mu^{2}>\omega^{2}\;.
  \end{equation}
To integrate the system of equations (\ref{eq:m}, \ref{eq:sigma},  \ref{eq:kg-mol}) outward from some small $r_{\rm min}$, the three constants $\omega$, $\sigma_0$ and $\psi_0$ must be provided. However,  a look at the equations shows that these constants can be reduced to two by scaling $\sigma$ and $\omega$ as
\begin{eqnarray}
    \sigma(r) \rightarrow \frac{\sigma(r)}{\sigma_0}\; , \qquad {\rm and} \qquad 
    \omega \rightarrow \frac{\omega}{\sigma_0}\; .
\end{eqnarray}
with this, the unknown constant $\sigma_0$ parametrizes the outer boundary condition. 
With this choice, only two constants, $\psi_0$ and $\omega$, must be specified at the inner boundary.

\section{Results}
\label{sec:results}

\subsection{Hayward Spacetime}
In the electro-vacuum limit, that is, where the scalar field is zero $\psi(r)=0$ in the whole domain, the equation for $\sigma$  Eq. \eqref{eq:sigma} and the equation for $m$ Eq. \eqref{eq:m-rho} can be solved separately once the regularity conditions at the origin $r=0$ is imposed and asymptotically flatness is guaranteed integration of \eqref{eq:sigma} under these conditions gives that $\sigma(r) = 1$. \\
A direct integration of Eq. $\eqref{eq:m-rho}$ with $\rho_{\Psi}= 0$ produces:
\begin{eqnarray}
    m(r) = \frac{2Q^2 r^3}{(2\beta Q^2)^{1/4}r^3+2\beta Q^2} = \frac{m_0 r^3}{r^3+\ell^3}  \; . 
\end{eqnarray}
the second equality is used to compare with the known Hayward spacetime \cite{Hayward:2005gi} and
where the relation between the charge $Q$ and the electrodynamics parameter $\beta$ 
with the Hayward length-scale $\ell$ and mass $m_0$, is
\begin{eqnarray}\label{eq:hay-params}
       \ell^2 = Q(2 \beta)^{1/2} \;, \qquad m_0 = Q^{3/2} \left(\frac{8}{\beta}\right)^{1/4} \; .
\end{eqnarray}
Recalling that the metric coefficient $ N(r) = 1- \frac{2m(r)}{r}$,
the location of the horizons in the Hayward spacetime is determined by the condition $N(r) = 0$. The nature of the roots is determined by a limiting length $\ell_m$,
for which the spacetime may be
a black hole with two horizons if $\ell < \ell_m$, an extreme black hole if $\ell = \ell_m$ and a spacetime without horizons if $\ell > \ell_m$
with
\begin{eqnarray}\label{eq:limit-lm}
    \frac{\ell_m}{m_0} = \frac{(32)^{1/3}}{3} = 1.05827... \; . 
\end{eqnarray}

\begin{figure}[ht!]
	\centering
	\includegraphics[width=0.6\linewidth]{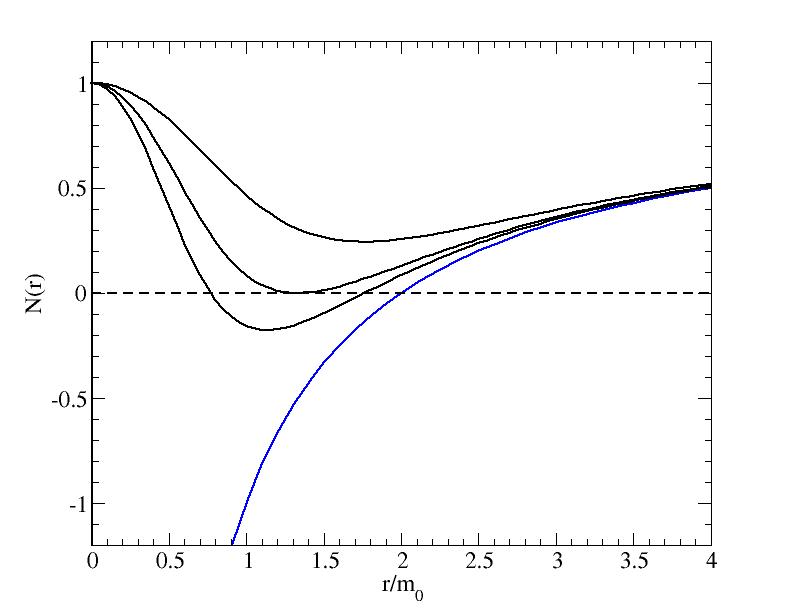}
	\caption{Metric coefficient $N(r)$ for the Hayward spacetime. Depending on the value of $\ell/m_0$ the spacetime may contain, one, two or no horizons. The location of the horizons is given by the roots of $N(r)$. The extreme value $\ell_m/{m_0}=1.05827$ produces the extreme black hole with only one horizon at $r_{\rm ext,bh}=4m_0/3$. The blue line corresponds to Schwarzschild spacetime (with $\ell=0$).}
	\label{fig:metricF}
\end{figure}

\begin{figure}[ht!]
	\centering
	\includegraphics[width=0.6\linewidth]{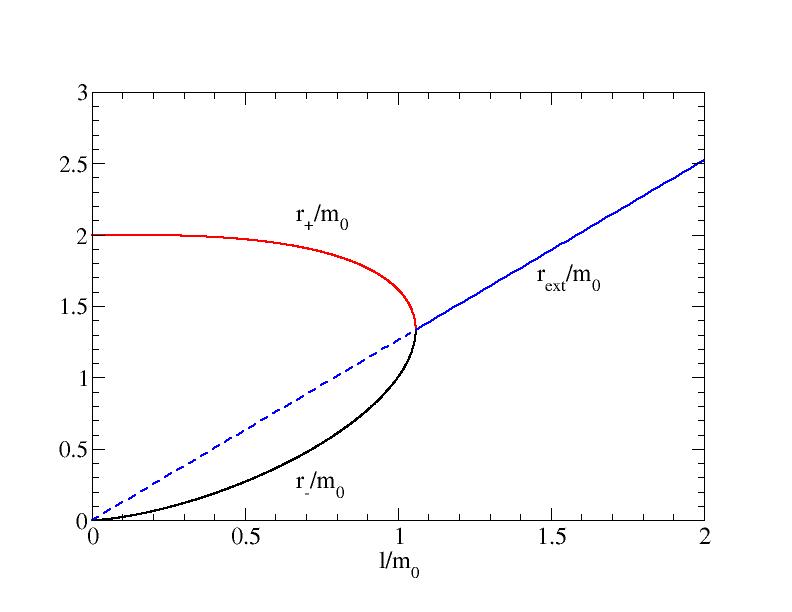}
	\caption{The radius of the event horizon $r_+$ and inner horizon $r_{-}$. The line  $r_{\rm ext}=2^{1/3}\ell/m_0$, marks the location of the minimum of the metric function $N(r)$. For non extreme black holes, the minimum is located between the horizons (dashed line). In the extreme case ($r_{\rm ext,bh}=4m_0/3$) the three radii coincide.}
	\label{fig:horizonss}
\end{figure}

The outer $r_{+}$ and inner horizon $r_{-}$ exist for values $0\leq \ell \leq \ell_m$ \cite{Chiba:2017nml} and are given by 
\begin{eqnarray}\label{eq:horizon}
 r_{+}/m_0 = \frac{2}{3}+\frac{4}{3}\cos\left[ \frac{1}{3}\cos^{-1}\left( 1- \frac{27\ell^3}{16m_0^3} \right) \right] \ , \qquad
   r_{-}/m_0 = \frac{2}{3}-\frac{4}{3}\cos\left[ \frac{1}{3}\cos^{-1}\left( 1- \frac{27\ell^3}{16m_0^3} \right) +\frac{\pi}{3}\right]
  \ .
\end{eqnarray}
Figure \ref{fig:metricF} shows the metric coefficient $N(r)$ of the Hayward spacetime, for some values of $\ell/m_0$, representing the three possible cases for the spacetime, a black hole with two horizons a extreme black hole with one horizon or a spacetime with no horizons.
For comparison, the Schwarzschild limit ($\ell = 0$) is also shown. Notice that the value of $r_+$ is always smaller than the Shwarzschild radius $r=2m_0$.
Figure \ref{fig:horizonss} shows the inner (Cauchy), outer (event) horizon as a function of the ratio $\ell/m_0$.
The figure also displays the 
radius $r_{\rm ext}$ for which the minimum of the metric function $N(r)$ is attained.

The event horizon is a decreasing function starting from $r_+=2m_0$, the Schwarzschild limit with $\ell=0$, whereas the inner horizon is an increasing function. The three radii coincide in the extreme case  $\ell_m/{m_0}=1.05827$ where 
$r_{+}=r_{-}=r_{\rm ext,bh}=4m_0/3$.
As the value of $\ell/m_0$ increases, the roots of $N(r)$ become complex and the spacetime has no horizons. However, there is still a localized concentration of energy. We characterize the \emph{size} of the resulting object with the radius $r_{\rm ext}$ (which satisfies the equation $N'(r_{\rm ext})=0$; $\Rightarrow$ $r_{\rm ext}=2^{1/3}\ell/m_0$ ). 
This radius will become useful when comparing the size of this object with the effective size of boson stars presented in the following section. 
Fig. \ref{fig:horizonss}  shows that for 
values 
$\ell>\ell_m$ the radius $r_{\rm ext}$ grows linearly with  $\ell$ as stated above. In other terms,
the size of a horizonless Hayward spacetime is bounded from below by $r_{\rm ext,bh}$.

\begin{figure}[ht!]
	\centering
	\includegraphics[width=0.48\linewidth]{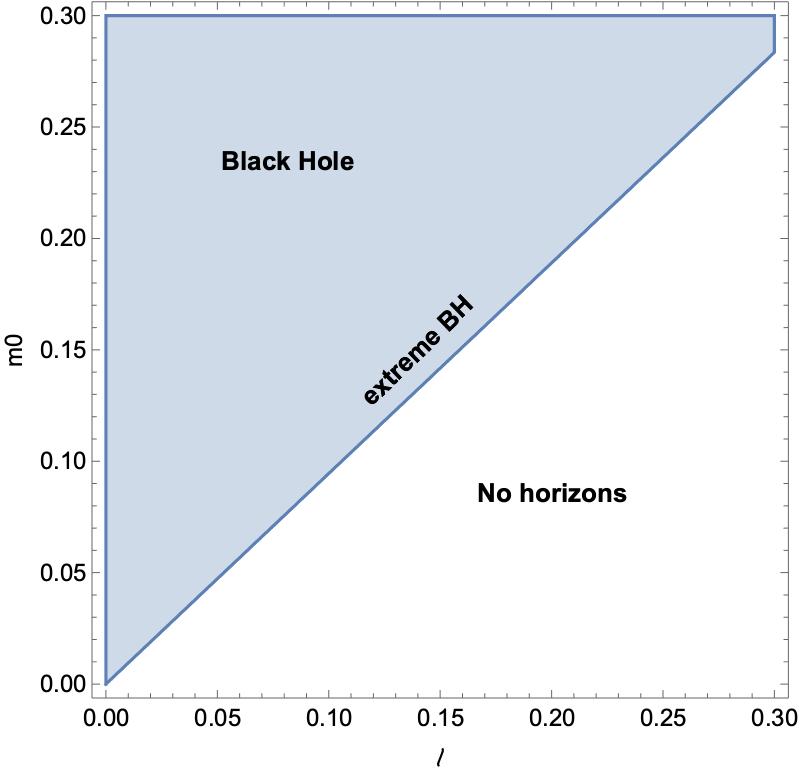}
	\caption{The existence region for black holes and a spacetime with no horizons. The diagonal line is given by expression $\ell/{m_0}=1.05827$ as indicated in Eq.~\eqref{eq:limit-lm}.}
	\label{fig:limit-lm}
\end{figure}

Figure \ref{fig:limit-lm} shows the plane $\ell$ \emph{vs.}\,$m_0$. The  region of existence of black holes has been shaded. The separating line is the representation of Eq. 
\eqref{eq:limit-lm}.   
For very small values of $\ell$ and larger masses $m_0$, the resulting spacetime is regular with no horizons.

\subsection{Boson stars}

In order to solve the system (\ref{eq:m}, \ref{eq:sigma} and \ref{eq:kg-mol}) numerically, we use  dimensionless quantities defined as
\begin{eqnarray}
    r \rightarrow\mu r\; , \quad
     \omega \rightarrow \frac{\omega}{\mu}\; , \quad
    m \rightarrow\mu m\; , \quad
        Q \rightarrow\mu Q\; , \quad
            \beta \rightarrow\mu^2 \beta\; .
\end{eqnarray}
This scaling, leaves the equations invariant under the values of the mass parameter, allowing the solutions to be scaled with $\mu$. 

We solve the system for static solutions using a shooting method. We begin by choosing values for $\psi_0$ and $\omega$. With these we get the values of the fields at some small $r_{\rm min}$ through the conditions \eqref{eq:initial}.
We then numerically integrate the solution outward from $r_{\rm min}$ at some large $r_{\rm max}$. 
The equations are solved by using a standard
Runge-Kutta ODE.
In general, the integrated solution at large $r$ will not
satisfy the outer boundary condition $\psi\rightarrow 0$.
The shooting method consist on varying the constant $\omega$ until it does. Once the outer boundary condition are satisfied, we have found a static solution.

In this work, we focus exclusively on ground state solutions, which are characterized by the absence of nodes in the radial profile of $\psi$ for 
$r>0$ \cite{Seidel:1990jh}. To construct Hayward boson star configurations, we began with known solutions for mini-boson stars and a fixed value of 
$\beta$ and $Q$ then
gradually increasing the value of $\psi(0)$ to generate a family of solutions. 

Given a static solution, the asymptotic value of $m(r)$ is the Arnowitt-Desser-Misner (ADM) mass $M$, which is approximated by evaluating $m(r)$ at the last grid point $r_{\rm out}$ of the computational domain.

Although boson stars extend to infinity and hence do not have a surface, an effective size $R_{99}$ 
is defined for them as the radius of the object
in which $99\%$ of the total mass is enclosed. 

After obtaining solutions for larger 
values of $\psi_0$, we also investigated configurations with larger values of $Q$,
to ensure a consistent exploration of the solution space.
However, as was pointed out in \cite{Yue:2023sep} the value of the charge can not grow without limit since 
there is a critical value in which solutions do not exist.
In the following section we will provide an explanation for the existence of this bound in the value of the charge.

\begin{figure}[ht!]
	\centering
\includegraphics[width=0.6\linewidth]{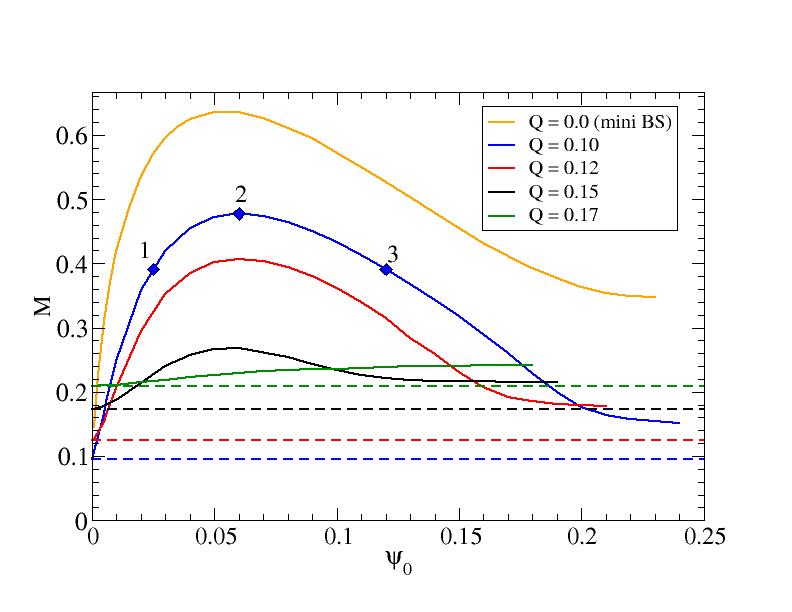}
	\caption{Mass \(M\) as a function of the central scalar field value \(\psi_0\) for configurations with different values of \(Q\). The horizontal dashed lines are the values of $m_0$, the mass of the equivalent electrovacum Hayward spacetime. Configurations marked with 1,2,3 will be analized below.}
	\label{fig:mass-psi0}
\end{figure}

Figure \ref{fig:mass-psi0} displays the space of static solutions. The curves shown are constructed from a series of discrete points, each point representing a static configuration.
Each of these configurations is characterized by a central value of the field $\psi_0$, an associated frequency 
$\omega$, a total mass $M$, and an effective radius $R_{99}$.
In the figure we present the equilibrium configurations for some representative values of $Q=$ $\{0.10,\, 0.12, \, 0.15, \, 0.17\}$ and $\beta=0.1$
 showing the total mass as a function of $\psi_0$. The mass increases monotonically with the central value of $\psi$ up to a maximum point, after which it begins to decrease mirroring the known behavior of standard mini-boson stars (with $Q = 0$). In the mini boson star case the maximum value of the mass $M_{\rm max}$ separates stable from unstable configurations \cite{Seidel:1990jh}. 
 
As stated above, for each value of $Q$ (with a fixed $\beta$) there is a value of $m_0$ and $\ell$ that yields an electro-vacuum Hayward spacetime.
The horizontal dashed lines in Fig. \ref{fig:mass-psi0} correspond to the equivalent electro-vacumm Hayward mass $m_0$, for each value of $Q$ as given in Table \ref{tab:mass-max} constructed with the help of Eq. \eqref{eq:hay-params}.
Since the energy density of the scalar field is strictly positive [see Eq. \eqref{eq:energy-density}], its presence contributes to add mass to the system relative to the electro-vacuum spacetime. As a result, the total mass of the HBS is always greater than that of its electro-vacuum counterpart, and is in fact bounded from below by the corresponding value of $m_0$.

\begin{figure}[ht!]
	\centering
\includegraphics[width=0.6\linewidth]{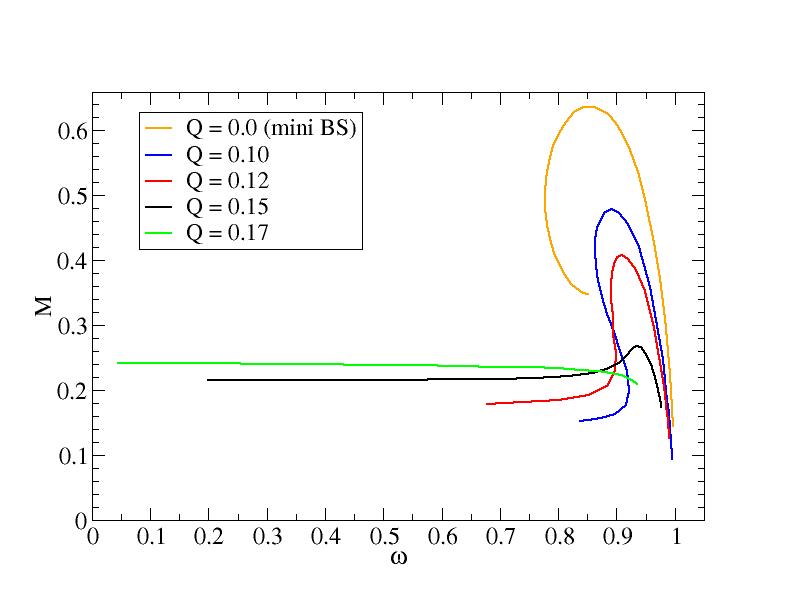}
	\caption{ Mass $M$, as a function of the frequency $\omega$, for the same configurations of 
    Fig. \ref{fig:mass-psi0}.
    The spiraling behavior of mini-boson starts is lost progressively for larger values of $Q$ in the strong gravity regime.
     }
	\label{fig:mass-w}
\end{figure}

The curves in Fig. \ref{fig:mass-w} are another presentation of the space of static 
solutions, showing the mass as a function of the frequency.
We see that static solutions exist in the range $0< \omega/\mu<1$ and that as the value of $Q$ increases, the frequency of the solutions goes to zero, this is the limit of \emph{frozen boson stars} 
described in Refs. \cite{Chen:2024bfj,Wang:2023tdz}. 
Furthermore, the characteristic spiral structure observed in mini-boson star solutions \cite{Herdeiro:2015gia} gradually diminishes as the charge 
$Q$ increases, eventually disappearing for sufficiently large values of 
$Q$ (up to a maximum value). The plot also reveals another important result; the maximum mass attainable by any HyBS is lower than that of mini-boson stars, whose maximum mass is
$M_{\rm max}\simeq 0.633$. \cite{Balakrishna:1997ej}
	 
\begin{figure}[ht!]
	\centering
\includegraphics[width=0.6\linewidth]{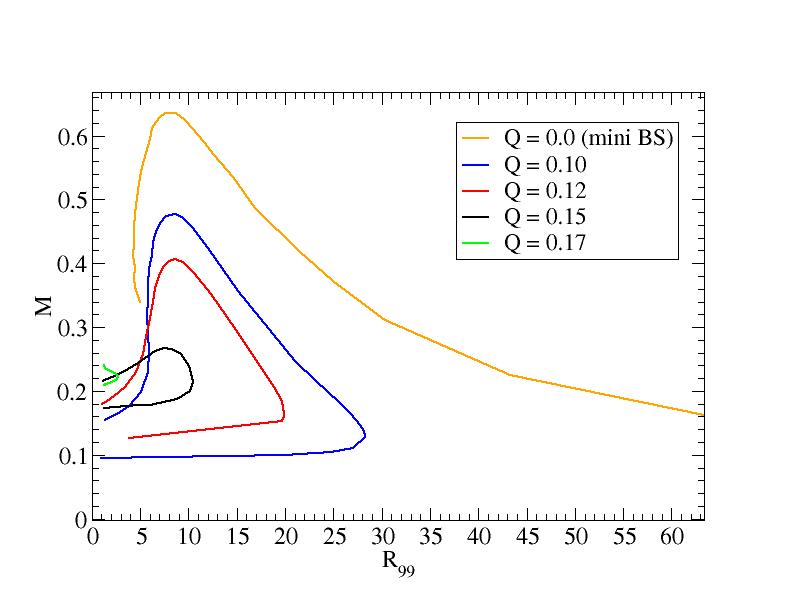}
	\caption{Mass \(M\) as a function of \(R_{99}\)  for different values of \(Q\). For $Q\neq 0$, configurations with less scalar field, start from the lower part of the curve with a small radius, as the contribution of the field increases, the mass grows slowly until a certain radius and then grows faster up to a maximum value but with a smaller size.}
	\label{fig:mass-r99}
\end{figure}

In Fig. \ref{fig:mass-r99} we show a plot for the total mass as a function of the effective radius $R_{99}$,
the mass increases monotonically with $R_{99}$ up to a maximum value, and then decreases. 
For mini-boson stars,  $M$ decreases monotonically with $R_{99}$, 
and beyond the maximum, configurations become diluted in the sense that they have smaller mass and larger size
(this corresponds to the Newtonian gravity regime).
For $Q\neq 0$ a couple comments are in order. 
First, after the maximum is reached and as $R_{99}$ increases, the curves turn towards smaller values of $R_{99}$, thus, unlike miniboson stars they can not attain an arbitrary large size.
Second, after the turn, the effective size of the HBS decreases to the value $r_{\rm ext}$ of the electrovacuum Hayward spacetime while their mass tends to the constant value, $m_0$.
which is consistent with the fact that the limit of HyBS of $\psi\rightarrow 0$ is the Hayward spacetime, once the parameters $Q$ and $\beta$ are provided.

Table \ref{tab:mass-max} contains the mass $M_{\rm max}$, effective size $R_{99}$, and the inverse of compactness $R_{99}/M_{\rm max}$, of the maximum mass configurations for some representative values of Q.
The table 
also shows
the mass $m_0$, effective length $\ell$, and size $r_{\rm ext}$ of the 
equivalent electro vacuum Hayward spacetime  obtained from relations \eqref{eq:hay-params}. The ratios $\ell/m_0$ and 
$r_{\rm ext}/m_{0}$ are also provided.


One can see that, as the value of $Q$ increases, the ratio $\ell/m_0$ of the configurations decreases towards the value of the extreme Hayward 
BH with $\ell_0/m_0=1.05827$ [see Eq. \eqref{eq:limit-lm}].
This limit value would correspond to a configuration with $Q\simeq 0.21$. However, HyBS with this and higher charges does not exist.
This expectation is confirmed by looking at the
radius $r_{\rm ext}$ in Table \ref{tab:mass-max}. Moving to higher values of $Q$ reveals that $r_{\rm ext}$ tends to the radius of the extreme Hayward BH $r_{\rm ext, bh}=1.33 m_0$.

\begin{figure}[ht!]
	\centering	
\includegraphics[width=0.45\linewidth]{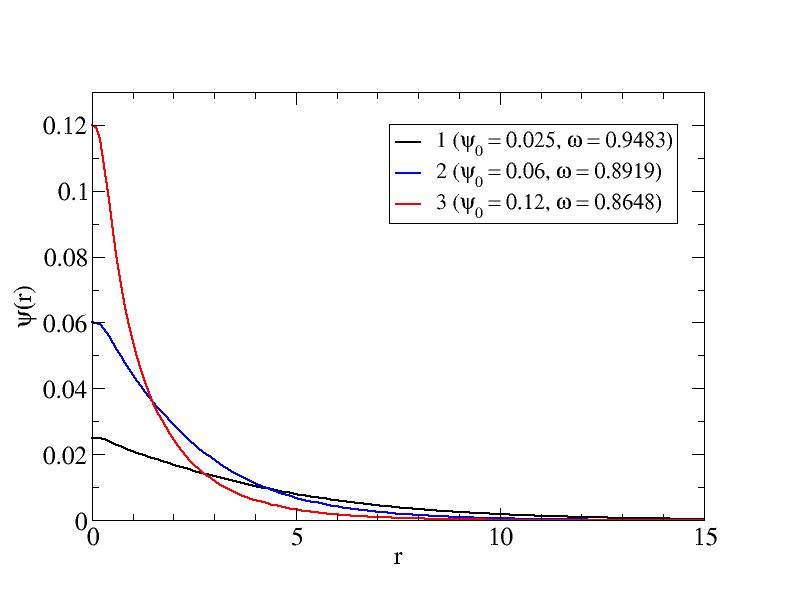}
\includegraphics[width=0.45\linewidth]{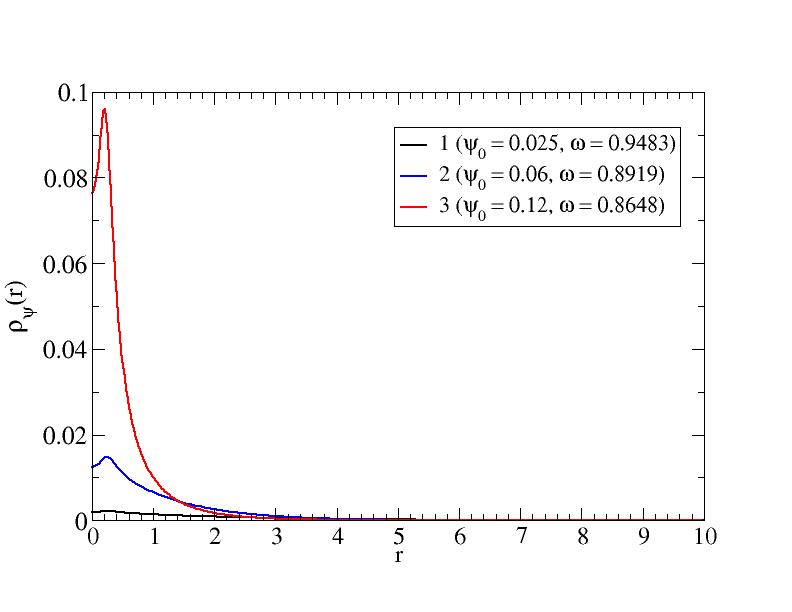}		\caption{Scalar field profile $\psi$, and scalar field density $\rho_\psi$, for the configurations labeled with 1, 2 and 3 in Fig. (\ref{fig:mass-psi0}).}
	\label{fig:psi-rhoRCL-Q0.1}
\end{figure}

Regarding properties of individual configurations,
Figure \ref{fig:psi-rhoRCL-Q0.1} displays the scalar field profile $\psi(r)$ and the scalar energy density  $\rho_\psi(r)$, for solutions marked with 1, 2 and 3 in Fig. \ref{fig:mass-psi0}.
The figure displays only the scalar field component of the energy density.
Solution 2 corresponds to the one with maximum mass
$M_{\rm max}\simeq0.477$ for $Q=0.1$. Configurations 1 and 3 have nearly the same total mass $M\simeq0.391$.
The maximum of the scalar field profile is at the origin and then decreases as $r$ increases. Figure \ref{fig:psi-rhoRCL-Q0.1} also shows that the scalar profile becomes sharper in the near origin region for larger values of $\psi_0$.

Remarkably, unlike mini boson stars in spherical symmetry,
the maximum of the scalar energy density for HyBS is not located at the origin.
For solution marked with 1, the density is more spread in space compared with solution 2 while the density for solution 3 has a larger value than the others.
Since the maximum amplitude of the scalar field is located the origin but the maximum in the density is not there, we conclude that the shift in the location of the maximum is being caused (through gravity) by the presence of the electromagnetic contribution $\rho_{\rm EM}$.

\begin{figure}[ht!]
	\centering	
\includegraphics[width=0.45\linewidth]{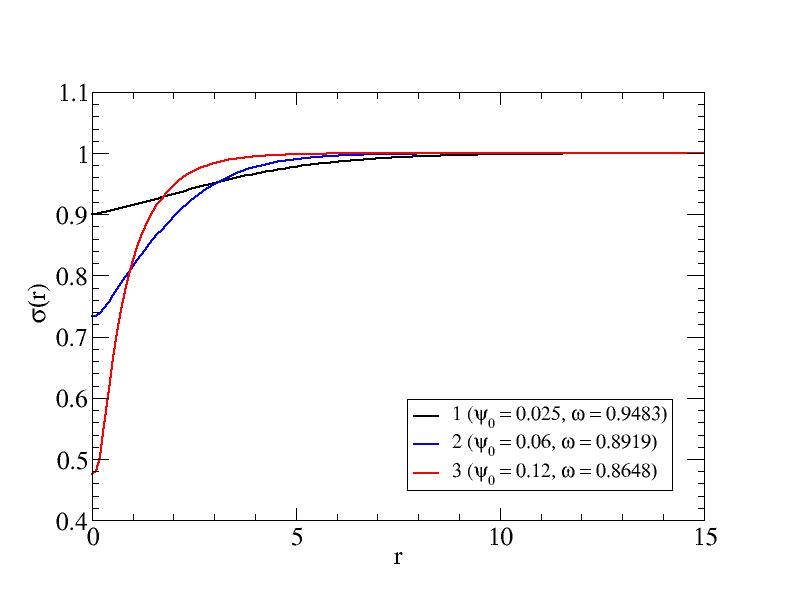}
\includegraphics[width=0.45\linewidth]{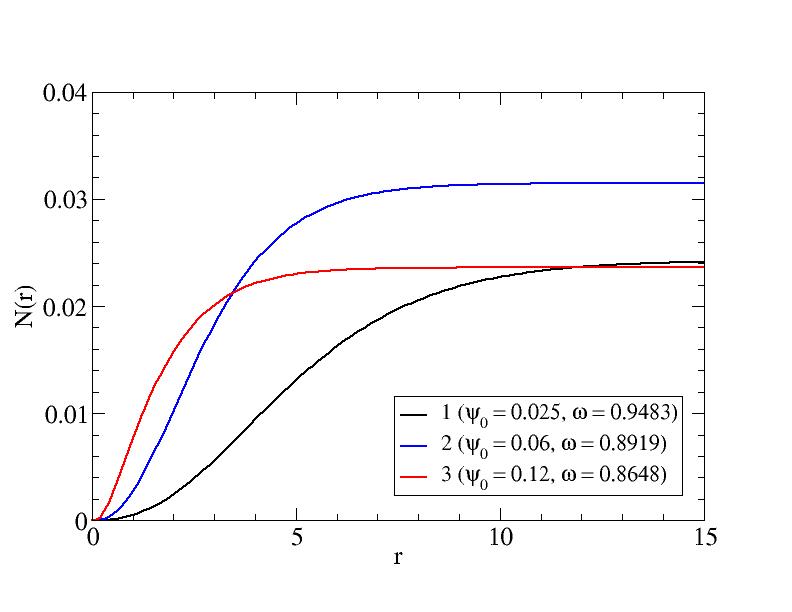}
	\caption{Metric coefficients $\sigma$ and $N$, as a function of $r$ for the configurations displayed with labels 1, 2 and 3 in Fig. \ref{fig:mass-psi0}.}
	\label{fig:sigma-NRCL-Q0.1}
\end{figure}

The metric coefficients $\sigma$ and $N$ as functions of $r$ for the configurations 1, 2 and 3 in Fig. \ref{fig:mass-psi0} are shown in Fig. \ref{fig:sigma-NRCL-Q0.1}. In all cases, the spacetime is asymptotically flat, that is, $\sigma \rightarrow 1$
and $N\rightarrow const.$ as $r\rightarrow \infty$.

\begin{figure}[ht!]
	\centering	
\includegraphics[width=0.6\linewidth]{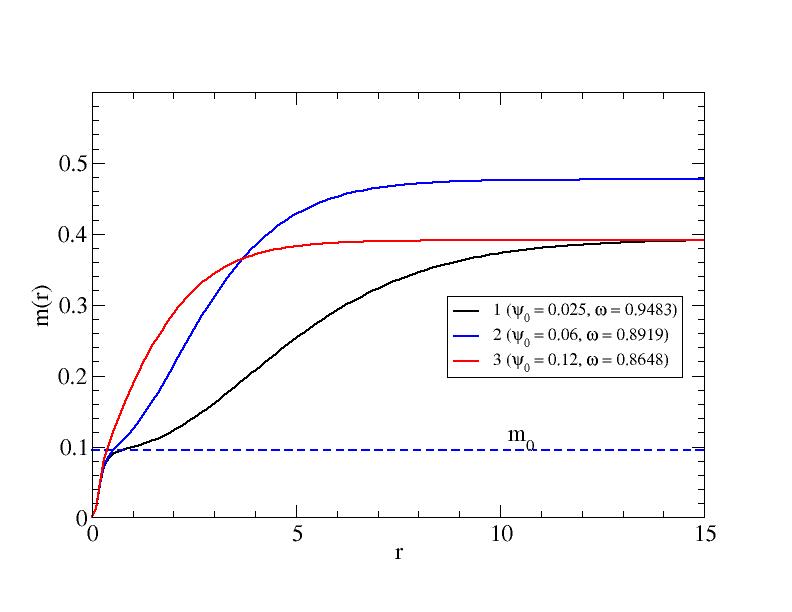}
	\caption{Mass profile as a function of $r$ for configurations with labels 1, 2 and 3 in Fig. (\ref{fig:mass-psi0}). The horizontal dashed line marks the value of $m_0$ that corresponds to the Hayward spacetime with parameters
    $Q=0.1$ and $\beta = 0.1$ ($\ell=0.21147$ and $m_0=0.094574$ respectively.)
    }
	\label{fig:massRCL-Q0.1}
\end{figure}

Figure \ref{fig:massRCL-Q0.1} shows the mass function for configurations 1, 2 and 3.
Asymptotically, curves 1 and 3 converge to the same value, that is, the ADM mass of the HyBS.
Close to the origin, a slight bending is present in curves 1 and 2 being more noticeable in the first where the contribution of the scalar field is smaller. This occurs because, in the region near origin, the electromagnetic contribution to the spacetime is larger than the one of the scalar field, and, as one moves outward, the scalar field contribution is dominant, producing a more massive object. 
As one moves in the solution space (curve Fig. \ref{fig:mass-psi0}) towards larger values of $\psi_0$, as for configuration 3, the contribution of the scalar fields dominates from the very beginning. Remarkably, the mass of the configurations does not grows without bound but reaches a maximum (configuration 2) and then decreases.

%

%

\begin{table}[ht]

\caption{The maximum mass attained by the Hayward boson stars $M_{\mathrm{max}}$. 
For an extreme Hayward black hole $r_{\rm ext}=4/3m_0$ and thus $C_{\rm ext}= m_0/r_{\rm ext}=3/4=0.75$ }
\centering{}%
\begin{tabular}{cccc|ccccc}
\hline 
$Q$&  $M_{\rm max}$& $R_{99}$& 
$R_{99}/M_{\rm max}$  
&$m_{0}$ & $\ell$ & $\ell/m_0$& 
$r_{\rm ext}$
& 
$r_{\rm ext}/m_0$
\tabularnewline
\hline 
\hline 
$0$ &$0.633$ &  $7.855$ & 
12.1409 & 
$-$ & $-$  & $-$ &  $-$ &  $-$ 
\tabularnewline
$0.1$ & $0.477$ 
& $8.602$ & 
17.9987 & 
$0.0945$ & $0.2114$  & $2.2360$ &  $2.8172$ &  
29.7903
\tabularnewline
$0.12$ &$0.407$ 
& $8.598$
&
21.1123 & 
$0.1243$ &  $0.2316$ & $1.8633$ &  $2.3477$ &  
18.8843
\tabularnewline
$0.15$ &  $0.267$  & $7.499$ 
& 
28.0287 & 
$0.1737$ &  $0.2590$ & $1.4907$ &  $1.8781$ &  
10.8101
\tabularnewline
$0.17$ & $0.241$  & $1.216$
&
4.6482 & 
$0.2096$ &  $0.2757$ & $1.3153$ &  $1.6572$ &  
7.9056
\tabularnewline
\hline 
\end{tabular}\label{tab:mass-max}
\end{table}

\section{Discussion and final remarks}
\label{sec:conclusions}


Once the parameter of the theory  $\beta$ and the charge $Q$ are fixed, if they correspond in electrovacum to a regular horizonles spacetime, 
it becomes possible to sustain a bounded, self-gravitating, and regular scalar field configuration or a HyBS. 
Conversely, when the chosen parameters in electrovacum yied to horizons, self-gravitating scalar field configurations cannot be sustained.

\begin{figure}[ht!]
	\centering
	\includegraphics[width=0.47\linewidth]{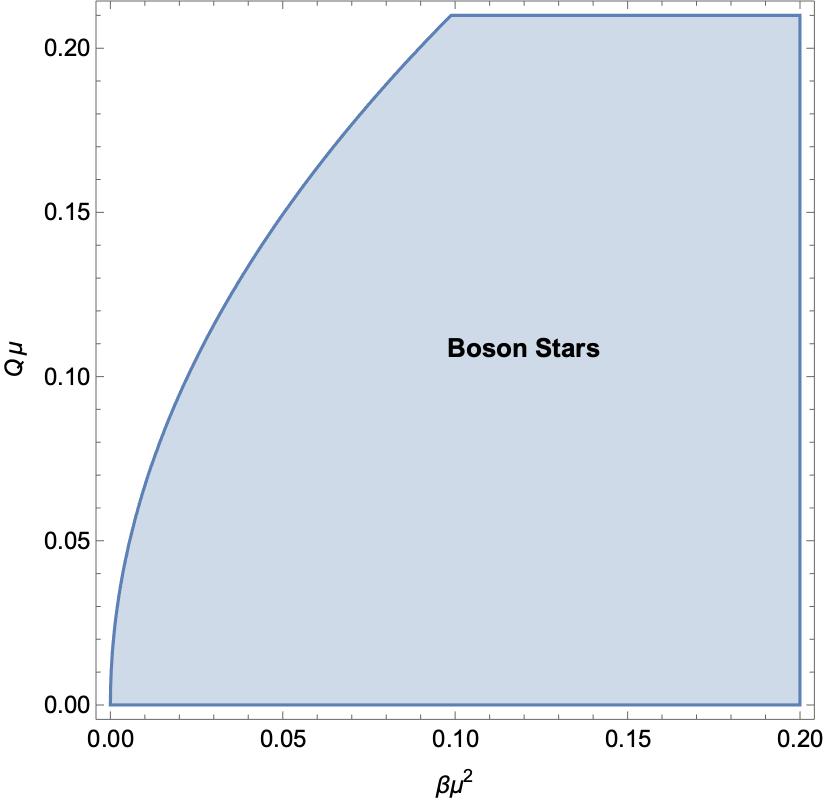}
	\caption{
    The shaded region marks the region where boson stars exist in the presence of an electromagnetic field sourcing a Hayward spacetime for the combination of parameters $\beta$ and $Q$. 
    }
	\label{fig:limit-boson}
\end{figure}

The region of the parameter space defined by $\beta$
and 
$Q$, where HyBS solutions exist, is shown in Fig. \ref{fig:limit-boson}.
The condition of existence for HyBS can be found using the correspondence between the Hayward parameters with $\beta$
and 
$Q$, and is given by [See \eqref{eq:limit-lm}] 
\begin{eqnarray}
    \ell/m_0 = \frac{1}{Q}\sqrt{\frac{\beta}{2}}> 1.05827 \; , \qquad \Rightarrow \qquad \frac{\sqrt{\beta}}{Q}> 1.49661 \; .
\end{eqnarray}


Different results in the literature establish the impossibility of having Shwarzschild or Reissner-Nordstrom BHs with scalar hair \cite{Mayo:1996mv,Martinez:2004nb,Hertog:2006rr,Gao:2021ubl}.
Some of these results can be derived from 
integral relations that can also be applied in the case of a Hayward Black Hole.
In the following, we use one of these arguments
to provide further insight into the existence of the BH solutions coupled with a scalar field.

One starts assuming a non-extremal Hayward BH. 
Then re-write the scalar field equation \eqref{eq:psi} in the form
\begin{eqnarray}
(Nr^2\sigma \psi')'=
    r^2\sigma\left(\mu^2-\frac{\omega^2}{N\sigma}\right)\psi \; .
\end{eqnarray}
After integrating between the event horizon and infinity, one finds the following
\begin{eqnarray}\label{eq:int-nohair}
    (Nr^2\sigma \psi')\Huge|_{r=r_{+}}^{\infty}
= \int_{r_{+}}^{\infty}r^2\sigma\left(\mu^2-\frac{\omega^2}{N\sigma}\right)\psi \;dr \; .
\end{eqnarray}
Since $N(r_+)=0$ and 
$\psi'(\infty)=0$
the \emph{l.h.s.}  of \eqref{eq:int-nohair}
vanishes. Furthermore, 
the term in parentheses in the \emph{r.h.s.} of \eqref{eq:int-nohair} is strictly positive outside the horizon, thus, solutions necessarily imply that the scalar field must vanish outside the horizon.


In conclusion, we have shown that the existence of HyBSs is guaranteed by a combination between the free parameter of the theory $\beta$, and the magnetic charge that sources the spacetime $Q$.
We found that the threshold values for the existence of HyBS is related to the existence of horizons in the equivalent electro-vacuum Hayward spacetime.
If the combination of the parameters of the Hayward spacetime, namely the ADM mass and length $\ell$, is such that, an event horizon appears, then the spacetime can not support a stationary scalar field
configuration. This result can be stated as a no hair theorem. \textit{Hayward black holes can not support self gravitating non-charged scalar field hair}.

Understanding the dynamical stability of these stars is important for evaluating their viability and physical plausibility as alternative models to black holes.
Such analysis  will be addressed in a future work.


\acknowledgments
This work was partially supported by 
DGAPA-UNAM through grant IN110523 
and by the programme HORIZON-MSCA2021-SE-01 Grant No. NewFun-FiCO101086251.


\bibliography{ref}


\end{document}